\documentclass[]{article}
\usepackage{amsmath}
\usepackage{amssymb}
\usepackage{subfigure}
\usepackage{epsfig}
\title{Authenticated tree parity machine key exchange}

\author{Markus Volkmer and Andr\'e Schaumburg\\
Hamburg University of Science and Technology\\
Department of Computer Engineering VI\\
D-21073 Hamburg, Germany\\
\texttt{\{markus.volkmer,a.schaumburg\}@tuhh.de}\\
}
\date{}
\begin{document}
\maketitle
\begin{abstract}
The synchronisation of Tree Parity Machines (TPMs), has proven to
provide a valuable alternative concept for secure symmetric key exchange. Yet, from a cryptographer's point of view, authentication is at least
as important as a secure exchange of keys. Adding an authentication
via hashing e.g.~is straightforward but with no relation to {\em Neural Cryptography}.
We consequently formulate an authenticated key exchange within this
concept. Another alternative, integrating a Zero-Knowledge protocol
into the synchronisation, is also presented. 
A Man-In-The-Middle attack and even all currently known attacks, that are based on using identically structured TPMs and synchronisation as well, can
so be averted. This in turn has practical consequences on using the
trajectory in weight space. Both suggestions have the advantage of not affecting the previously observed physics of this interacting system at all.
\end{abstract}
\section{Introduction}
The symmetric key exchange method based on the fast synchronisation of two
identically structured Tree Parity Machines (TPMs) was proposed by
Kanter and Kinzel \cite{KKK02}. 
Their exchange protocol is realized implicitly by a mutual adaptation process
between two parties $A$ and $B$, not involving large numbers and
methods from number theory \cite{RKKK02a}. 

Making sure, that the two parties involved are also allowed to perform
this protocol is the cryptographic process of {\em (entity) authentication}. In the area of
cryptography, authentication is an important step still before key exchange or even the
en-/decryption of information with an exchanged secret key \cite{MOV01}. Adding classical authentication e.g. via hashing to
the Neural Cryptography concept 
is straightforward but is not embedded into the concept itself. We think it is thus desirable to formulate an authentication
concept from within Neural Cryptography, based on the original TPM
synchronisation principle and keeping the practical advantage of not
operating on large numbers.

We first briefly recapitulate the parallel-weights version, in which weights are
identical in both TPMs after synchronisation, using
hebbian learning and the so-called {\em bit package} variant of the protocol \cite{KKK02}. 
The anti-parallel-weights version, using anti-hebbian learning and
leading to inverted weights at the other party, can be considered for
our purpose as well but is omitted for brevity.
The notation $A/B$ denotes equivalent operations for the parties $A$ and $B$. A single $A$ or $B$ denotes an operation which is
specific to one of the parties.

\begin{figure}[ht!]
 \begin{center}
 \subfigure[]{\includegraphics[width=6.5cm]{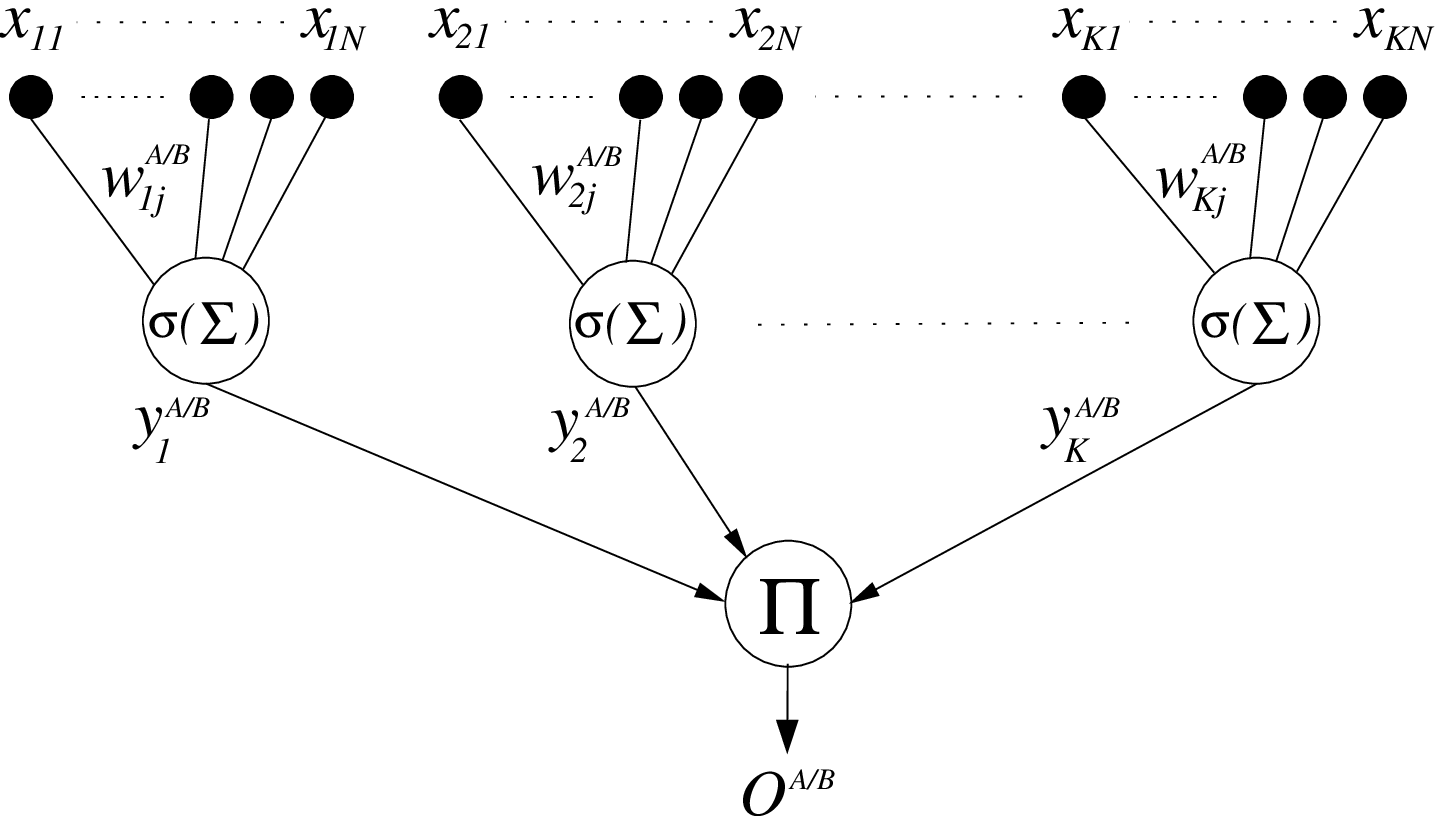}}\hspace{0.2cm}
 \subfigure[]{\includegraphics[width=1.5cm]{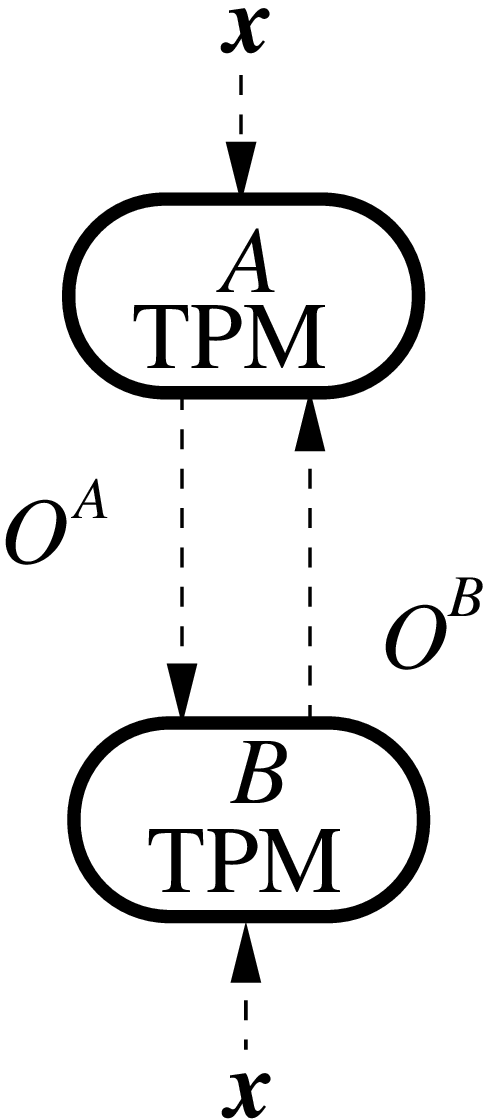}}
 \caption{(a) The tree parity machine (TPM) generates a single output -- the parity of the outputs of the hidden units. (b) For mutual
 learning, outputs on commonly given inputs are exchanged between the
 two parties $A$ and $B$.}
 \label{PM_fig}
 \end{center}
 \end{figure}

The TPM consists of $K$ hidden units ($1\leq k\leq K$)
in a single hidden-layer with non-overlapping inputs and a single unit
in the output-layer. The particular tree structure 
 has binary inputs, discrete weights and a single
 binary output as depicted in \mbox{Figure \ref{PM_fig}a}.

Each hidden unit $k$ receives different $N$ inputs $x_{kj}(t)$ ($1\leq j\leq N$), leading
to an input field of size $K\cdot N$. The vector-components
$x_{kj}(t)\in\{-1,1\}$ are random variables with zero mean and unit
variance. They can e.g. be coded as bits generated by a {\em Linear Feedback Shift
Register} (LFSR) as pseudo-random number generator.
The output $O^{\scriptscriptstyle{A/B}}(t)\in\{-1,1\}$, given
bounded weights \mbox{$w_{kj}^{\scriptscriptstyle{A/B}}(t)\in[-L,L]\subseteq \mathbb{Z}$}
(from input unit $j$ to hidden unit $k$) and common pseudo-random inputs $x_{kj}(t)$, is calculated by a parity function of the
signs of summations:
\begin{eqnarray}\label{out}
O^{\scriptscriptstyle{A/B}}(t)&=&\prod_{k=1}^{K}y^{\scriptscriptstyle{A/B}}_k(t)\nonumber\\
&=&
\prod_{k=1}^{K}\sigma\!\left(\sum_{j=1}^{N}w_{kj}^{\scriptscriptstyle{A/B}}(t)\
x_{kj}(t)\right)\ .
\end{eqnarray}
$\sigma$ is a 
sign-function.

Parties $A$ and $B$ start with an individual randomly generated initial
weight vector $w^{\scriptscriptstyle{A/B}}(t_{\scriptscriptstyle{0}})$ -- their
secret. After a set of $b>1$ presented inputs, where $b$ denotes the size of the
bit package, the corresponding $b$ TPM outputs
$O^{\scriptscriptstyle{A/B}}(t)$ are exchanged over the public
channel in one package (see \mbox{Figure \ref{PM_fig}b}).
The $b$ sequences of hidden states
$y^{\scriptscriptstyle{A/B}}_k(t)\in\{-1,1\}$ are stored for the subsequent learning process.
A hebbian learning rule is applied to adapt the weights, using
the $b$ outputs and $b$ sequences of hidden states:
\begin{equation}\label{learn}
w_{kj}^{\scriptscriptstyle{A/B}}(t):=w_{kj}^{\scriptscriptstyle{A/B}}(t-1)+O^{\scriptscriptstyle{A/B}}(t)\
x_{kj}(t)
\end{equation}
They are changed according to Equation \ref{learn} only on an agreement
$O^{\scriptscriptstyle{A}}(t)=O^{\scriptscriptstyle{B}}(t)$ on the parties' outputs. Furthermore, only weights of 
those hidden units are changed, that agree with this output, i.e.~if~$O^{\scriptscriptstyle{A/B}}(t)=y^{\scriptscriptstyle{A/B}}_k(t)$.
Updated weights are bound to stay in the maximum range
\mbox{$[-L,L]\subseteq \mathbb{Z}$} by reflection onto the boundary
values. 

Synchrony is achieved when both parties have learned to produce
each others outputs. They remain synchronised (see \mbox{Equation~\ref{learn}}) and
continue to produce the same outputs on every commonly given input.
This effect in particular leads to common weight-vectors
$w^{\scriptscriptstyle{A/B}}(t)$ in both TPMs in each following iteration. These weights have never been communicated
between the two parties and can be used as a common time-dependent key for
encryption and decryption respectively. 

A test for synchrony can of course not practically be defined by checking whether weights in both
nets have become identical. One rather tests on successive equal outputs in a
sufficiently large number of iterations $t_{\scriptscriptstyle{min}}$, 
such that equal outputs by chance are excluded. 
 \begin{equation}\label{synccrit}
 \forall t\in[t',\cdots,t'+t_{min}]:\
 O^{\scriptscriptstyle{A}}(t)=O^{\scriptscriptstyle{B}}(t)\ .
 \end{equation}
The synchronisation time was found to be finite for discrete weights. It is almost
independent on $N$ and scales with $\ln N$ for very large $N$, even in the
thermodynamic limit $N\rightarrow \infty$. Furthermore, it is proportional to $L^2$ \cite{MPKK02b}. 
 Our investigations confirmed that the average synchronisation time
 is distributed and peaked around 400 for the parameters given in \cite{KKK02}.
 The number of bits required to achieve synchronisation is lower than
 the size of the key \cite{KKK02,MPKK02b}. 
Secret key agreement based on interaction over a public insecure
channel is also discussed under information theoretic aspects by
 Maurer \cite{Maurer93a}, also with regard to unconditional security.

\section{Authentication through secret common inputs}
In the original key exchange protocol, the structure of the network,
the involved computations producing the output
$O^{\scriptscriptstyle{A/B}}(t)$ \mbox{(Equation~\ref{out})}, the adaptation-rule
(Equation~\ref{learn}
) and especially the common inputs $x_{kj}(t)$ are public. The only secrets involved are
the different initial weights
$w_{kj}^{\scriptscriptstyle{A/B}}(t_{\scriptscriptstyle{0}})$ of the
two parties. If they were not secret, the resulting keys could simply be calculated (by an
adversary), because all further computations are completely deterministic.  

An elegant solution to include authentication into the neural key
exchange protocol comes from the observation, that two parties $A$ and $B$ which do not have the same
input vectors 
\begin{equation}
\forall t:\ x^{\scriptscriptstyle{A}}(t)\not=
 x^{\scriptscriptstyle{B}}(t) 
\end{equation}
cannot synchronise. 
Remember, that the aim of the two-party-system is
to learn each others outputs on commonly given inputs. Given different
inputs, the two parties are trying to learn completely different
relations (two different nonlinear mappings) between inputs
$x^{\scriptscriptstyle{A/B}}(t)$ and outputs $O^{\scriptscriptstyle{A/B}}(t)$. Consequently, when the two parties
do not synchronise, there also will not be time-dependent equal
weights $w^{\scriptscriptstyle{A/B}}(t)$ and thus no exchange of a key. 
This again is exactly the service one would want to restrict only to authorised parties employing an
explicit authentication. 

We experimentally investigated the development of normalised sum of absolute differences
$d(w^{\scriptscriptstyle{A}}(t),w^{\scriptscriptstyle{B}}(t))\in [0,1]$ over
time for different offsets 
\begin{equation}
\forall t\!:\
 x^{\scriptscriptstyle{A}}(t)=x^{\scriptscriptstyle{B}}(t+\Delta),\
 \Delta \in\mathbb{N}
\end{equation}
 in the (pseudo-random) input-list and for completely different
input-lists. The first situation represents an attacker, who has a
different initialisation of his pseudo-random number generator. The
second situation is typical for an attacker with incomplete or even
completely differently generated inputs.
\begin{figure}[htb]
\begin{center}
\includegraphics[width=7cm]{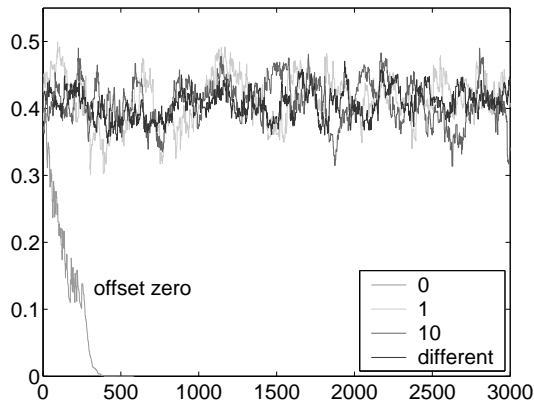}\label{synch}
\caption{Distance $d$ 
 vs. the number of
exchanged bits (\mbox{iterations $t$}) for offset zero
(successful authentication), offsets one and ten, as well as for completely different
inputs.}
\end{center}
\end{figure}
One can observe in Figure~\ref{synch}, that the distance between two parties that do not
possess the same inputs remains fluctuating within a certain limited
range around $0.4$ and never decreases towards zero. We also investigated different
offsets with the same qualitative outcome. Two parties with completely different
inputs (although not realistic given a concrete and publicly known LFSR as pseudo random
number generator) show the same qualitative behaviour. Considering the number of repulsive and attractive steps, one can
constitute, that on average there must be as many repulsive as
attractive steps for such a behaviour (cf. \cite{RKSK04}). Two parties
having the same inputs (offset zero) soon decrease their distance and synchronise. 

Another test was performed with identical inputs but by imposing a certain percentage of equally distributed `noise'
on the communicated outputs of one party. It allows to demonstrate the
importance of common inputs for the synchronisation process. If such a noise would appear
only in a certain period, the system would still synchronise but with a
delay of roughly the length of the noisy period plus the time used up
for unsuccessful synchronisation before the noisy period, which is thus
not the interesting case. 
\begin{figure}[ht]
\begin{center}
\includegraphics[width=7cm]{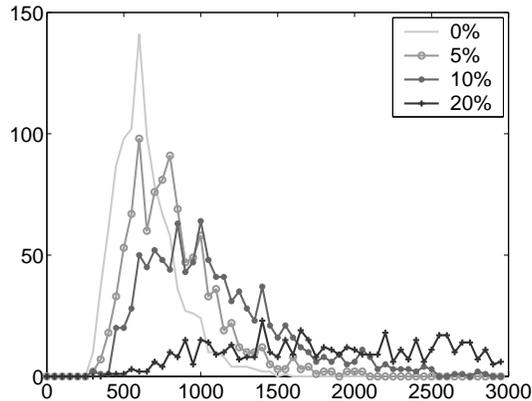}\label{synchnoise}
\caption{Peaks of the histogram (average over 1000 runs) of the iterations necessary for synchronisation for
different percentages of noise on the communicated output bits of one
party. The curves for one and two percent noise were omitted, as they almost match with the zero percent curve.}
\end{center}
\end{figure}

As can be seen in Figure~\ref{synchnoise}, the
distribution of synchronisation times is flattened and biased towards
longer times for increasing noise. Surprisingly, the system can still synchronise even
with highly noisy communication. Obviously, the (coordinated) inputs
basically determine the synchronisation. The average synchronisation time is
of course increased as is the probability for a late synchronisation.

A superficial explanation of the observed behaviour is, that the principle is based on mutual learning from common
 inputs and thus on principle cannot work with differing inputs. 
More concretely, the random walks with reflecting boundaries performed by the weights in the
iterative process now make uncorrelated moves and moves in the wrong direction
(cf. \cite{RKSK04,RRK04}). 
Two corresponding components $w_{kj}^{\scriptscriptstyle{A}}(t)$ and
$w_{kj}^{\scriptscriptstyle{B}}(t)$ now receive a different random
component $x_{kj}(t)$ of their (differing) input vectors (cf. Equation~\ref{out}). The distance between the components is thus no longer successively
reduced to zero after each bounding operation 
and the two parties diverge. 

The non-synchronisation in the case of no
common inputs, therefore enables us to incorporate authentication by
keeping the common (pseudo-random) inputs
$x^{\scriptscriptstyle{A/B}}(t)$ secret between the two parties in
addition to their individual secret (random) initial weights
$w^{\scriptscriptstyle{A/B}}(t_{\scriptscriptstyle{0}})$. There are
$2^{KN}-1$ possible common inputs as second initial secrets,
which is a large enough practical amount for the parameters as chosen
in \cite{KKK02} that makes brute force attacks computationally very
expensive. Even more, a Man-In-The-Middle attack and all other
currently known attacks \cite{KMS02asia,KKSKM04} using TPMs are averted on principal by such an
authentication. It is important to note, that such a second secret does not represent any
principal disadvantage, because a basic common information is always
also necessary in other authentication protocols (cf.
\cite{MOV01}). 

As opposed to asymmetric approaches in which a third party that can be trusted issues 
a second public key, in this symmetric approach a second secret 
information is necessary for authentication, with the advantage of not 
requiring a central authority. Using an asymmetric public-key
authentication like e.g. in the Fiat-Shamir
authentication scheme, a trusted center selects and publishes an
RSA-like modulus, which is the second common (but public) information in addition to the
private key. Therefore security is partly transferred to a third trusted party.
\section{Embedding a Zero-Knowledge protocol}
Although we have authentication already given the second secret
described above, we make another suggestion explicitly incorporating a
{\em Zero-Knowledge} (ZK)
protocol (see e.g. \cite{MOV01}). It also requires a (second) secret but formally does not
require the non-synchronisation in case of differing inputs. Although this may seem redundant at first glance, it allows to demonstrate
how the two (already) interactive protocols can be merged and allow a
quicker authentication at the cost of an only statistical and thresholded
secure authentication. 
ZK mechanisms generally allow to split a protocol into an iterative process of
relatively light transactions, instead of a single (heavy)
transmission. 
Typically such a principle depends on random numbers in some way. The security that can be achieved is probabilistic,
i.e. depending on the number of interactions, but security can always be
increased beyond some acceptable variable security threshold. 

Again we take the inputs of the TPM as a second common secret.
The probability of an input vector $x^{\scriptscriptstyle{A/B}}(t)$ having a
particular parity $p\in \{0,1\}$ is 0.5. This parity will now be used directly as an
output bit $O^{\scriptscriptstyle{A/B}}(t)$ for an
authentication step. 
The probability of both parties having the same output
bit upon a given input at any given time $t$ is
\begin{equation}
P(O^{\scriptscriptstyle{A}}(t)=p=O^{\scriptscriptstyle{B}}(t))=1/2.
\end{equation}
Given a number $n\ (1\leq n \leq \alpha)$ of pure authentication
steps, in which one transmits the parity of the corresponding input
vector as output $O^{\scriptscriptstyle{A/B}}$ directly, the
probability that the two parties subsequently produce the same output
$n$ times (and thus are likely to have the same $n$ inputs) decreases exponentially with $n$
\begin{equation}
P(O^{\scriptscriptstyle{A}}(n)=O^{\scriptscriptstyle{B}}(n))=1/2^n\
;\forall n\ .
\end{equation}
Consequently, in order to have a statistical security of $\epsilon \in [0,1[$ one has to pick $n=\alpha$ authentication steps such that 
\begin{equation}
1-1/2^\alpha\geq \epsilon
\end{equation}
which can be calculated in advance as 
\begin{equation}
\alpha=\bigg\lceil\log_2\left(\frac{1}{1-\epsilon}\right)\bigg\rceil\ .
\end{equation}
One achieves a statistical security $\epsilon = 0.9999$ (i.e. $99.9999$ \%) with $\alpha = 14$, for example.
The synchronisation time for the ZK variant thus increases by $\alpha$
authentication-steps depending on the required level of security $\epsilon$. 

The questions arises, when to perform those authentication steps and what happens in the case of a synchronisation
earlier than authentication, which is possible due to the distribution
of synchronisation times? One obviously has to pick those entries in the input list used for
authentication only such that the security threshold will be reached soon enough with a certain probability. 
This can be achieved by selecting a certain bit sub-pattern in the
input vector. Inputs are equally distributed by definition and thus 
the last say $m$ bit are also equally distributed. One can thus select those
entries that possess a defined bit sub-pattern (e.g. `0101' for $m=4$). The probability of such
a fixed bit sub-pattern of $m$ bit to occur is $1/2^m$, because each bit has
a certain fixed value with a probability of $0.5$ and the individual
bits occur independently from the LFSR.
Thus for four bit, on average every 16th input would be used for
authentication. When this sub-pattern occurs, one performs an authentication step in
transmitting the parity of the corresponding input vector directly as output $O^{\scriptscriptstyle{A/B}}(n)$.
This will (definitely) only happen at the other party (and with the
same output!) if it has the same inputs. Having successfully performed
$\alpha$ authentication-steps, one commences with the synchronisation
and key exchange. 

Such an authentication does not influence the
learning process at all, which transfers all behaviour of the TPM
synchronisation to this extended principle.
Due to the fact that the inputs are secret, an attacker
cannot know when exactly such an authentication step is happening. 
This e.g. would not be the case, if one would reserve the first iterations only for authentication. An attacker could just record one session
and replay the authentication steps (using the recorded outputs) when performing his attack.

Let us elaborate on three important properties of a ZK protocol (cf. e.g. \cite{MOV01}) and see how they apply in the context of
proposed authentication principles:
\begin{enumerate}
\item {\em Completeness -- $A$ always succeeds in convincing $B$ if he knows
the common secret:} If $A$ knows the common secret in the form of
having the same inputs, he will always synchronise within a finite time (typically
around 400 iterations for the parameters used in \cite{KKK02}). In the case of the second authentication principle, $A$ will reach the
security threshold $\epsilon$ in the specified $\alpha$ authentication steps. Thus both protocols are complete.
\item {\em Soundness -- $A$ succeeds with (arbitrary) small probability if he does not know
the secret of $B$:} If $A$ does not know the common secret and has
different inputs, synchronisation will fail. The two parties will
always be driven apart again by the repulsive steps. He will thus
succeed with a probability of zero.
In the case of the explicit authentication principle, $A$ will not reach the
security threshold $\epsilon$ in the specified $\alpha$ authentication steps and will
be rejected. Thus both protocols are sound.
\item {\em Zero-Knowledge -- No information on the common secret is leaked at all while the
interactive protocol is performed.} This property can be attributed back to the lack of information in the transmitted
output bits (or Bit Packages). The only information transmitted
is the parities of unknown bit-strings. The same holds for the
parities of the inputs chosen (pseudo-randomly) only for authentication in the case of the explicit authentication principle. Again only the
parities of randomly generated input bit vectors are transmitted. An
attacker also cannot distinguish an authentication step from a
synchronisation step from observing the exchanged outputs. He thus does not know, whether
the currently observed output bit is used for either of the two
purposes if he does not know the second secret. Both protocols thus
possess the Zero-Knowledge property.
\end{enumerate}
Both suggestions for authentication could after all be viewed as
ZK protocols, one implicit and one explicit, due to their interactive questioning nature
that does not reveal information on the common secret. Furthermore,
any previous findings on the physics of the synchronisation of TPMs still apply.  
Obviously, the bit packaging variant of the protocol together with the
 ZK extension is a typical parallel interaction
 protocol (cf.~\cite{MOV01}). In such a parallel protocol, a number of
problems ($b$ outputs of party $A$) are posed
 an and a number of solutions ($b$ corresponding outputs of party $B$)
at a time are asked. This is generally
 used to reduce the number of interaction messages with a
 slow-response-time connection or low-bandwidth channel. 

The general trade-off in cryptography between available resources and the required level
of security also applies using the TPM principle. In many practical
embedded security solutions e.g. it is often admissive to provide a system safe enough for the
 particular application, and given certain attack scenarios. The TPM
principle extended with the proposed authentication is very attractive for such embedded applications
due to its hardware-friendly basic operations \cite{VW04,VWTC}.
\section{Consequences on using the weights' trajectory}\label{traj}
As mentioned in the introduction, once synchronous, the two parties
remain synchronised having identical weights in each following
iteration. This mode of operation was regarded potentially insecure
by the authors in \cite{KK02,K03} with respect to the possible attacks with
identical TPMs on the ongoing communication. 
We would like to comment on that with two basic considerations: 
\begin{enumerate}
\item When the two parties are synchronous they will also have the same
outputs in each iteration. Thus, one can as well turn off the communication,
because all following outputs will be identical anyway and thus do not need to be
communicated any longer. Each party can then simply apply the learning
rule (Equation~\ref{learn}) with its own
output. Consequently, staying in the trajectory does not automatically
represent a security weakness as stated in \cite{KK02,K03}. Only if a TPM attacker achieved to synchronise before or at the
same time as the two parties, he will have the keys from the trajectory. But the problem of a possible attack on the ongoing communication
can be avoided as described above.  
\item Given the herein proposed authentication refutes the currently
known attacks with TPMs on principle. An attacker with a TPM will not be successful
in synchronising, not even if the communication after synchronisation
goes on. This allows to securely exploit the full potential of
the trajectory.
\end{enumerate} 
In particular after having synchronised once, one can increase
the final key length by concatenating subsequently synchronised
`partial-keys' from the trajectory at the negligible cost of one or a
few further iterations, depending on the partial-keys length and the
desired final key length. Furthermore, one could even encrypt each
given data block to be transmitted securely with a separate key,
effectively yielding a {\em one-time pad} with a maximum length equal to the
length of the period (of the trajectory). In this case, even a less
sophisticated but low-cost encryption like simple XOR or LFSR becomes applicable.   

There are $2^{KNL}-1$ theoretically possible $K\cdot N\cdot L$ bit keys but the length of
the period (of the trajectory) has so far not been calculated. 
We also performed software simulations and did not find
two identical 612 bit keys in a million runs not using the trajectory. 
\section{Conclusion}\label{conc}
Two ways of establishing authentication from within the concept of
Neural Cryptography were presented. Next to the key establishment itself, such an authentication
is of primary interest in cryptography and its applications.
Using the common inputs as a second secret for authentication, we investigated the distance of the two parties' weight vectors for
different offset in their inputs and for completely different
inputs. No synchronisation appears, as expected. 
Another explicit authentication principle (based on the same underyling secret), naturally integrating a Zero-Knowledge protocol
into the already interactive key exchange concept was discussed and
concrete suggestions for its application were derived from probabilistic considerations.
It turns out that authentication is inherently provided by the
underlying synchronisation principle of Neural Cryptography. 

Above all, using authentication of this kind
averts all currently known attacks and a previously possible
Man-In-The-Middle attack, which assume the full knowledge on the
inputs to the TPMs. Any (non brute force) attack now needs to extract
information from the communicated outputs. Furthermore a (differential) power analysis on a concrete
software or hardware implementation could be tried, which is yet an
attack on a rather technical level. 
The outlined consequences of being able to securely stay in the
trajectory in weight space are of significant practical importance. 

It is thus our hope, that the discussion of this extraordinary key exchange
principle and related concepts (see e.g. \cite{Maurer93a}) will continue, within the physics community and also the cryptography community. 
\bibliographystyle{abbrv}

\begin{thebibliography}{10}

\bibitem{K03}
S.~Bornholdt and H.~Schuster, editors.
\newblock {\em Handbook of Graphs and Networks}, chapter Theory of interacting
  neural networks.
\newblock Wiley VCH, 2003.

\bibitem{KKK02}
I.~Kanter, W.~Kinzel, and E.~Kanter.
\newblock Secure exchange of information by synchronization of neural networks.
\newblock {\em Europhysics Letters}, 57(1):141--147, 2002.

\bibitem{KKSKM04}
I.~Kanter, W.~Kinzel, L.~Shacham, E.~Klein, and R.~Mislovaty.
\newblock Cooperating attackers in neural cryptography.
\newblock {\em Phys. Rev. E}, 69(066137), 2004.

\bibitem{KK02}
W.~Kinzel and I.~Kanter.
\newblock Interacting neural networks and cryptography.
\newblock In B.~Kramer, editor, {\em Advances in Solid State Physics},
  volume~42. Springer Verlag, 2002.

\bibitem{KMS02asia}
A.~Klimov, A.~Mityagin, and A.~Shamir.
\newblock Analysis of neural cryptography.
\newblock In {\em Advances in Cryptology -- \uppercase{AsiaCrypt} 2002}, volume
  2501 of {\em Lecture Notes in Computer Science}, pages 288--298. Springer
  Verlag, 2002.

\bibitem{Maurer93a}
U.~Maurer.
\newblock Secret key agreement by public discussion.
\newblock {\em IEEE Transactions on Information Theory}, 39(3):733--742, 1993.

\bibitem{MOV01}
A.~J. Menezes, P.~C. van Oorschot, and S.~A. Vanstone.
\newblock {\em Handbook of Applied Cryptography}.
\newblock CRC Press, 5th edition, 2001.

\bibitem{MPKK02b}
R.~Mislovaty, Y.~Perchenok, I.~Kanter, and W.~Kinzel.
\newblock Secure key-exchange protocol with an absence of injective functions.
\newblock {\em Phys. Rev. E}, 66(066102), 2002.

\bibitem{RKKK02a}
M.~Rosen-Zvi, E.~Klein, I.~Kanter, and W.~Kinzel.
\newblock Mutual learning in a tree parity machine and its application to
  cryptography.
\newblock {\em Phys. Rev. E.}, 66(066135), 2002.

\bibitem{RKSK04}
A.~Ruttor, W.~Kinzel, L.~Shacham, and I.~Kanter.
\newblock Neural cryptography with feedback.
\newblock {\em Phys. Rev. E}, 69, 2004.

\bibitem{RRK04}
A.~Ruttor, G.~Reents, and W.~Kinzel.
\newblock Synchronization of random walks with reflecting boundaries.
\newblock {\em J. Phys. A: Math. Gen.}, 37:8609--8618, 2004.

\bibitem{VW04}
M.~Volkmer and S.~Wallner.
\newblock A low-cost solution for frequent symmetric key exchange in ad-hoc
  networks.
\newblock In P.~Dadam and M.~Reichert, editors, {\em Proc. of the 2nd German
  Workshop on Mobile Ad-hoc Networks, WMAN 2004}, volume P-50 of {\em Lecture
  Notes in Informatics (LNI)}, pages 128--137, Ulm, Germany, Sep. 20th 2004.
  Bonner K\"ollen Verlag.

\bibitem{VWTC}
M.~Volkmer and S.~Wallner.
\newblock Tree parity machine rekeying architectures.
\newblock {\em IEEE Transactions on Computers}, 2004.
\newblock (to be published).

\end{thebibliography}

\end{document}